\documentstyle[11pt,epsfig,axodraw]{article} 
 \def\baselinestretch{1.3}
\newcommand{\ba}{\begin{array}}
\newcommand{\ea}{\end{array}}
\newcommand{\bd}{\begin{displaymath}}
\newcommand{\ed}{\end{displaymath}}
\newcommand{\be}{\begin{equation}}
\newcommand{\ee}{\end{equation}}
\newcommand{\bea}{\begin{eqnarray}}
\newcommand{\eea}{\end{eqnarray}}

\def\etal{ {\em et al.}~}

\def\q2 {q^2}

\def\er {\tilde{e}_R}
\def\sel {\tilde{e}}
\def\el {\tilde{e}_L}
 \def\N0{\widetilde \chi^0}

\def\mpT{p_T \hspace{-1em}/\;\:}

\def\go{\rightarrow}
\def\goes{\longrightarrow}
\begin{document}
\begin{flushright}
{\large TIFR/TH/00-15 \\April, 2000  \\ hep-ph/0003225}
\end{flushright}

\begin{center}
{\Large\bf Study of $R$-parity-violating supersymmetric signals\\ 
at an $e^-e^-$ collider }\\[20mm]
{\bf Dilip Kumar Ghosh\footnote{dghosh@theory.tifr.res.in} and Sourov Roy\footnote{sourov@theory.tifr.res.in}}\\[4mm]
{\em Department of Theoretical Physics, 
Tata Institute of Fundamental Research\\
Homi Bhabha Road \\
Mumbai - 400 005 \\
INDIA} 
\\[10mm]
\end{center}
\begin{abstract}
We study the pair production of right selectrons at a 500 GeV $e^-e^-$ 
collider followed by their decay into an electron and a lightest neutralino. 
This lightest neutralino decays into multifermion final states in the 
presence of $R$-parity-violating couplings. A detailed analysis of possible 
signals is performed for some important regions of the parameter space. The 
signals are essentially free from the standard model backgrounds. 
\end{abstract}

\vskip 1 true cm

\noindent
PACS NOS. : 12.60.Jv, 13.10.+q, 14.80.Ly

\newpage
\setcounter{footnote}{0}

\def\baselinestretch{1.8}
\section{Introduction}
Supersymmetry (SUSY), as one of the most attractive options beyond the 
standard model (SM), has been studied for the past few decades \cite{Rev_SUSY}.
From the theoretical point of view it offers a solution to the hierarchy 
problem. On the other hand, a lot of effort has been devoted to looking at 
the phenomenological consequences of SUSY both in low-energy 
processes and at high-energy colliders \cite{SUSY_search}. One of the 
candidates for a realistic model is the minimal supersymmetric extension of 
the SM. In the SM, it is not possible to write down interactions which violate 
baryon number ($B$) or lepton number ($L$). In the SUSY version of the SM, 
particle spectrum is doubled and baryon number and lepton number are 
assigned to the supermultiplets, leading to $\Delta B = 1$ or $\Delta L = 1$ 
interactions in the Lagrangian. In the minimal supersymmetric standard model 
(MSSM) it is assumed that $B$ and $L$ are conserved quantum numbers. This is 
ensured by imposing a discrete multiplicative symmetry called $R$ parity
\cite{Farrar-Fayet} which is defined as          

$$R = (-1)^{L + 3B + 2S}$$
where $S$ is the intrinsic spin of the particle.

It can be checked very easily that $R$ equals +1 for standard model particles 
and -1 for the superpartners. An immediate consequence of $R$-parity 
conservation is that the sparticles appear in pair at each interaction 
vertex. This leads to the fact that the lightest supersymmetric particle 
(LSP) is stable. The interactions of the LSP must be of weak 
strength because they are mediated by virtual sparticles which are known to 
be quite heavy (of the order of the electroweak scale). The most favorite 
candidate to become an LSP is the lightest neutralino and the search 
strategies for supersymmetry guided by the principle of $R$-parity 
conservation are to look for signals with large missing energy and momentum 
carried by an undetected neutralino~\cite{SUSY_search}. Also the LSP is a 
good candidate for the cold dark matter of the universe~\cite{SUSY_darkmatter}.

The conservation of $R$ parity, however, is not prompted by any strong 
theoretical reason, and theories where $R$ parity is violated through 
nonconservation of {\it either} $B$ {\it or} $L$ have been considered. Such 
scenarios can be studied by generalizing the MSSM superpotential to the 
following form~\cite{Rpv}:  

\begin{equation}
W = W_{MSSM} + W_{\not R},
\end{equation}
with
\begin{equation}
W_{MSSM} = {\mu} {\hat H}_1 {\hat H}_2 + h_{ij}^l {\hat L}_i {\hat
H}_1 {\hat E}_j^c
+ h_{ij}^d {\hat Q}_i {\hat H}_1 {\hat D}_j^c + h_{ij}^u {\hat Q}_i
{\hat H}_2 {\hat U}_j^c
\end{equation}
and
\begin{equation}
W_{\not R} = \lambda_{ijk} {\hat L}_i {\hat L}_j {\hat E}_k^c +
\lambda_{ijk}' {\hat L}_i {\hat Q}_j {\hat D}_k^c  
+ \epsilon_i {\hat L}_i {\hat H}_2 +
\lambda_{ijk}''{\hat U}_i^c {\hat D}_j^c {\hat D}_k^c .
\end{equation}
Here, ${\hat H}_1$, ${\hat H}_2$ are the ${\rm SU}(2)$ doublet Higgs 
superfields which give rise to the masses of down-type and up-type quark 
superfields, respectively, and $\hat L$ $(\hat Q)$ denote lepton (quark) 
doublet superfields. ${\hat E}^c$, ${\hat D}^c$, ${\hat U}^c$ are the singlet 
lepton and quark superfields. $i, j, k$ are the generational indices and we 
have suppressed the ${\rm SU}(2)$ and ${\rm SU}(3)$ indices. The 
$\lambda_{ijk}$ are antisymmetric in $i$ and $j$ while the $\lambda_{ijk}''$ 
are antisymmetric in $j$ and $k$. The first three terms in $W_{\not R}$ 
violate lepton number and the last term violates baryon number. It is obvious 
that both the $L$ and $B$ violating terms cannot be present if the proton is 
stable. In order to get a large proton lifetime ($\sim 10^{40}$ s) 
\cite{proton_decay} it is sufficient to demand that either $L$ or $B$ be 
violated which in turn breaks $R$ parity. $R$-parity violation leads to 
considerable changes in the phenomenology. The most important consequence is 
that the LSP can decay now. Also, the lightest neutralino need not be the LSP 
because it is no longer a stable particle. The lepton number and baryon number 
violating terms mentioned above have received a lot of attention and 
constraints have been derived on these new couplings from present experimental 
data~\cite{Dreiner}. Prospects of $R$-parity violation have been studied in 
the context of following present and future colliders: CERN $e^+e^-$
collider LEP, DESY $ep$ collider HERA, $p\bar p$ at Fermilab Tevatron,
CERN Large Hadron Collider (LHC), $e^+e^-$ and $e\gamma$ Next Linear
Collider (NLC) [\ref{Rpv_Collider}--\ref{Ghosh2}]. 
Here we investigate the signatures of $R$-parity breaking at future 
$e^-e^-$ linear colliders. Our aim is to study the pair production of 
right selectrons ($\er$) which will then decay into an electron and
a neutralino. Finally the neutralino will decay into multifermions through 
different $R$-parity-violating couplings.

In this paper we shall discuss the $R$ violation in three separate categories 
for the convenience of the analysis. We will consider, in turn, $W_{\not R}$
with either the $\lambda$, $\lambda'$, or $\lambda''$ terms existing in the 
superpotential at a time. The bilinear term $\epsilon_i {\hat L}_i {\hat H}_2$ 
is also a viable agent for $R$-parity breaking which can induce vacuum 
expectation values for the sneutrino fields and generates a tree-level mass 
for one of the neutrinos~\cite{Hall,Hemp}. This scenario has been studied by 
several authors in the context of recent results from SuperKamiokande (SK) 
data on atmospheric neutrinos~\cite{SuperK} and attempts have been made to 
find the correlation between the given pattern of neutrino masses and mixings 
and collider signatures of supersymmetry~\cite{Roy-Datta}. So far, no work has 
been reported which includes the study of $R$-parity violation through the 
bilinear term in the context of $e^-e^-$ colliders and we wish to discuss it 
in our future work which requires a separate analysis 
altogether~\cite{Ghosh-Roy}. 
 
The paper is organized as follows. In Sec. 2, we describe the 
physics goals of $e^-e^-$ colliders and their advantages and disadvantages 
from the point of view of a supersymmetry search. In Sec. 3 we will 
discuss the numerical results followed by our conclusions in Sec. 4.
             
\section{Search for supersymmetry at $e^-e^-$ collider}
As we know, the current $e^+e^-$ collider at LEP is at the verge of its 
closing. Apart from putting some lower bounds on different SUSY particles,
there has been no sign of new physics beyond the SM from LEP. Perhaps one can
hope to see some signals beyond the SM at run II of Tevatron and, of course, 
the LHC, but the clean environment of the next generation $e^+e^-$ linear 
collider will definitely complement the signatures from hadron colliders. 
Even if SUSY is discovered at LHC, NLC can be used as a machine for precision 
measurements for different SUSY parameters~\cite{peskin}. 

Before going on to the discussion of the supersymmetry search,  
let us first mention in brief the unique features of an 
$e^-e^-$ collider which establishes its importance in order to make 
model independent measurements at future high-energy physics 
experiments~\cite{Heusch}. First of all, it should be emphasized that at 
linear colliders the replacement of a beam of positrons with a beam of 
electrons can be achieved in a rather straightforward manner and can lead 
to the option of colliding electron beams.     	 

At $e^-e^-$ colliders, the initial energy is well known and both  
$e^-$ beams can be highly polarized so that the initial states are 
specified. The backgrounds are, in general, extremely suppressed and 
they can be reduced further with specific choices of the beam 
polarizations. However, the total electric charge $Q$ and total lepton number 
$L$ of $e^-e^-$ colliders forbid the pair production of most of 
the superpartners by virtue of total charge and lepton number conservation.
This is one disadvantage of $e^-e^-$ colliders where only selectrons can
be pair produced through the exchange of a Majorana neutralino in the 
${\em t}$ and $\em u$ channels~\cite{keung, frank} as shown in Fig. 1.
In contrast, at $e^+e^-$ colliders, selectron pair production occurs
through ${\em s}$-channel $\gamma$ and $Z$ exchange as well as through
${\em t}$-channel $\tilde \chi^0_i$ exchange. The interference between
the ${\em s}$- and ${\em t}$-channel diagrams is always destructive for
$\sqrt{s} > m_Z$ \cite{bartl}. In the $e^-e^-$ mode, since the 
${\em u}$-channel diagram is present along with the ${\em t}$-channel
diagram and the interference between them is constructive, the production 
cross section is always larger compared to the $e^+e^-$ mode. This cross 
section can be further increased by choosing the initial electron beam 
polarization properly. It has been shown \cite{frank} that the right selectron 
pair production cross section is largest for the right-polarized initial 
electron beam. This can be explained from the fact that in most of the MSSM 
parameter space the LSP is B-ino dominated, which has a larger coupling with 
$e_R \er$ compared to $e_L \el$. Furthermore, it turns out that the selectron 
pair production cross section for the unpolarized initial state is smaller 
than that of right-polarized electron beams.

Another important feature of the $e^- e^-$ collider is its behavior 
near threshold which shows a sharp rise in the selectron pair production 
cross section~\cite{feng}. This enables one to measure the selectron masses
very accurately. In contrast, at an $e^+e^-$ collider the threshold 
measurement is rather poor, which compels one to determine the $\sel$ mass 
(with an error of few GeV) from the measurement of the electron end point 
energy~\cite{feng}. The study of slepton flavor violation can also be done 
very effectively in an $e^- e^-$ collider.

\begin{figure}[hbt] 
\centerline{\epsfig{file=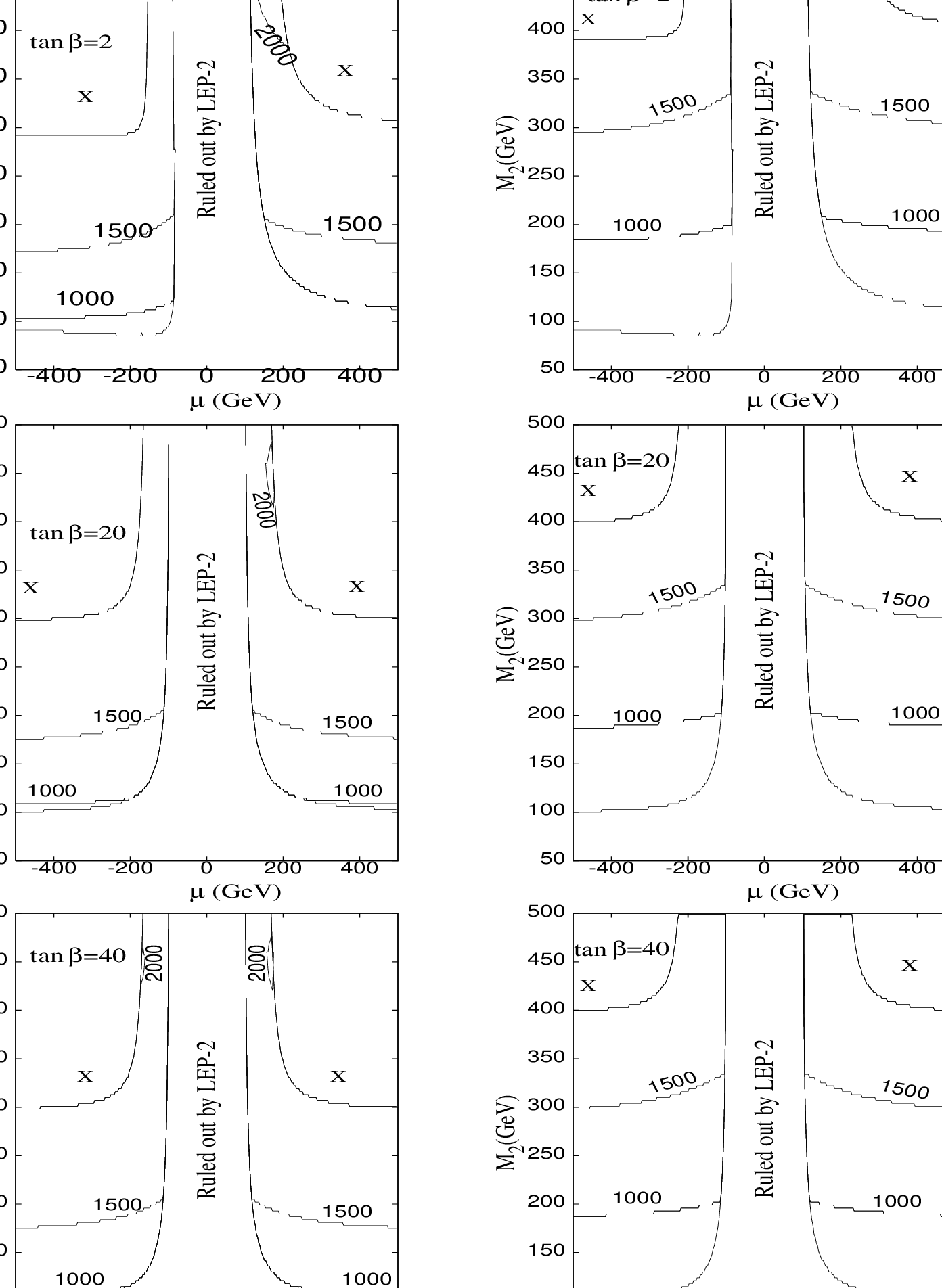,width=10cm}} 
\vspace{1.0in}
\caption{Contours of cross section (in ${\rm fb}$) for production 
of a pair of right selectrons at an $e^- e^-$ collider with 
right-polarized ($90\%$) electron beams. The left and right columns 
correspond to 150 GeV and 200 GeV right selectron masses.} 
\label{cont-prod}
\end{figure}

In Fig.~\ref{cont-prod}, we present contours for the cross section (in fb) for 
the production of $\er^- \er^-$ final states in the 
($\mu, M_2$) plane for $\tan\beta = 2, 20, 40$ and $\sqrt{s} = 500$~GeV. 
The mass of the right selectron is assumed to be 150 GeV for 
the plots in the left column and 200 GeV for the plots in the right 
column.

The explanation of the variation of cross section with the parameters 
which appear in the neutralino mass matrix, as shown in Fig.~\ref{cont-prod}, 
is as follows. The area {\em ruled out by LEP-2} represents the region 
which is disallowed by the chargino search at LEP-2 and corresponds to a 
mass of the lighter chargino ($\tilde \chi^\pm_1$) less than 98 GeV 
\cite{susylep}. This limit comes purely from kinematic considerations and 
does not depend on whether $R$ parity is conserved or violated. The area which 
is marked as $X$ in the figure is not allowed because here the selectrons 
become lighter than the LSP and hence the selectron decaying to the lightest 
neutralino is forbidden \footnote{In $R$-parity-violating models, it is 
also possible that the {\it selectron}, rather than the lightest neutralino, 
is the LSP, and can decay directly into two leptons (quarks) through the
$\lambda_{ijk} (\lambda^\prime_{ijk})$ couplings, respectively.}.
Since we are considering right selectron pair production, the contribution 
to the cross section comes mainly from the lightest neutralino which is 
dominated by a B-ino over a large part of the parameter space. Here we assume 
the grand unified theory (GUT) relationship between the ${\rm SU}(2)$ and 
${\rm U}(1)$ gaugino soft mass parameters $M_2$ and $M_1$, $M_1 = \frac{5}{3}
\tan^2\theta_W M_2$. As the value of $M_2$ increases, the lightest neutralino 
starts becoming more and more B-ino dominated and hence the strength of the 
$e_R-{\tilde e}_R-{\tilde \chi^0_1}$ coupling increases at the same time. 
Also, the amplitude in this case requires a ${\em t}$-channel neutralino mass 
insertion. These two effects combined together lead to an increase in
the cross section when $M_2$ is increased for a fixed value of 
$\mu$~\cite{keung}. This feature is evident from Fig. 2. For lower values of 
$\mu$, the B-ino component in the lightest neutralino starts decreasing which 
means a fall in the cross section and hence in order to get the same cross 
section the value of $M_2$ (consequently the value of $M_1$) must be increased. 
With the increase in selectron mass the available phase space reduces and 
in order to get the same cross section as in the left column one must go 
to higher values of $M_2$.  

The decay of right selectron yields following final state:
\bea
e^- e^- \goes \er^- \er^- \goes e^- e^-\N0_1\N0_1. 
\eea
This will give rise to two like-sign electrons and large $\mpT$ signature. 
This kind of a signal as shown in Eq. (4) and the relevant backgrounds 
have been well studied~\cite{frank}.   

In light of the above discussion, the next question which comes to 
mind is what could be the potential signatures at an  $e^- e^-$ 
collider when $R$ parity is violated. Recently, the effect of $R$ parity
violation has been studied for the production process 
$e^-_L e^-_R\goes \el^- \er^- $ \cite{umada}. In this work we will 
consider the pair production of right selectrons assuming $90\%$ 
right-polarized electron beams because of the larger cross section in this 
case. The subsequent analysis will not depend on the choice of initial 
electron polarization. As we will see in the following section, since the 
lightest neutralino will decay, it will lead to multilepton final states 
with missing energy almost free from standard model backgrounds. However, 
right selectrons can also decay into heavier neutralino states, if it is 
allowed by kinematics. In that case, the cascade decays of heavier neutralino 
will produce more complex signals. For the simplicity of our analysis, we
will not consider such decay patterns here. 

\section{Decay of $\N0_1$ and associated signals }
In this section we will discuss the possible signatures arising from 
the decay of the LSP through different $R$-parity-violating interactions, 
through sfermion (sleptons and squarks) exchange diagrams. These Feynman 
diagrams and the amplitudes can be found in the literature~\cite{Ghosh2,
Gondolo}. Here, we make the assumption that of all the couplings which violate 
$R$ parity, only one is dominant at a time, which is motivated from the 
fact that in the SM top quark Yukawa coupling is much larger than the others. 
Furthermore, we assume these couplings to be much smaller than the gauge 
couplings, though we require them to be large enough to make the LSP 
decay inside the detector. A generic $R$-parity-violating coupling should 
be larger than $10^{-5}$ to satisfy the above 
requirement~\cite{Dawson,Dreiner}. In our subsequent analysis we take these 
couplings in the range $10^{-1}$ - $10^{-2}$. 
If the $R$-parity-violating operator is of the type $LLE^c$, the final states
will have two charged leptons and a neutrino. The flavor of these leptons are
determined by the type of $\lambda_{ijk}$ coupling. 
If the $R$-parity-violating operator is of the type $LQD^c$, the final sates
will have either one charged lepton or a neutrino associated with two quarks.
Finally in the presence of baryon number violating coupling $U^cD^cD^c$, the
final state will have three quarks. Throughout this analysis we assume 250 GeV
left-slepton mass [sneutrino mass is related to left-slepton mass through 
the ${\rm SU(2)}$ relation] and 500 GeV squark mass. All the squarks have been 
assumed to be degenerate in mass. In our parton level Monte Carlo analysis we
treat quarks/partons as jets, and the direction of jets is same as that of 
the initial quarks/partons. We impose following selection criteria for these 
leptons and jets:
\bea
p_T^\ell > 5~{\rm GeV},~~|\eta_\ell| < 3,\\
~~p_T^{j}> 15~{\rm GeV},~~|\eta_{j}| < 3. 
\eea
We merge two jets into a single jet if their angular separation 
$\Delta R_{jj} < 0.7$, where 
${(\Delta R_{jj})}^2 \equiv {(\Delta \eta)_{jj}}^2 + {(\Delta
\phi_{jj})}^2$,
$\Delta\eta_{jj}$ and $\Delta\phi_{jj}$ being the difference of 
pseudorapidities and azimuthal angles, respectively, corresponding to two 
jets. The lepton is isolated from a jet if $\Delta R_{jl} > 0.4$, where 
$\Delta R_{jl}$ is defined in the same way as above. 

\subsection{Signals from $\lambda$-type couplings}
Let us now discuss the signals which can be looked for when $R$ parity is
violated through the terms of the type $\lambda ~L L E^c$. The
pair-produced LSPs from the decay of the two right selectrons 
will lead to the final state consisting of $e^-e^- +4\ell^\pm + \mpT$.
The flavor of the leptons coming from the neutralino
decay will depend on the particular type of coupling involved. For example,
$\lambda_{123}$ coupling gives
\bea
\N0_1 \goes \nu_e \mu^-\tau^+,~e^-\nu_\mu \tau^+,~\bar\nu_e \mu^+\tau^-,
~e^+\bar\nu_\mu \tau^-, 
\eea
with equal probabilities. Here, for simplicity we have considered a 
common value for all $\lambda$-type couplings taken to be 
$0.07 (m_{\tilde e}/100~{\rm GeV})$, close to the existing indirect bounds
relevant for most of those couplings. In order to tag the lepton flavor 
one must multiply the signal cross section with the efficiency of the 
corresponding lepton flavor identification. 

Since there are two neutrinos in the final states, reconstructing the 
mass of the LSP in such a case is not possible. However, the kind of 
final state mentioned above is spectacular in the sense that it is free 
from standard model background and permits easy detection at a 500 GeV 
$e^- e^-$ collider.
 
\begin{figure}[hbt]
\centerline{\epsfig{file=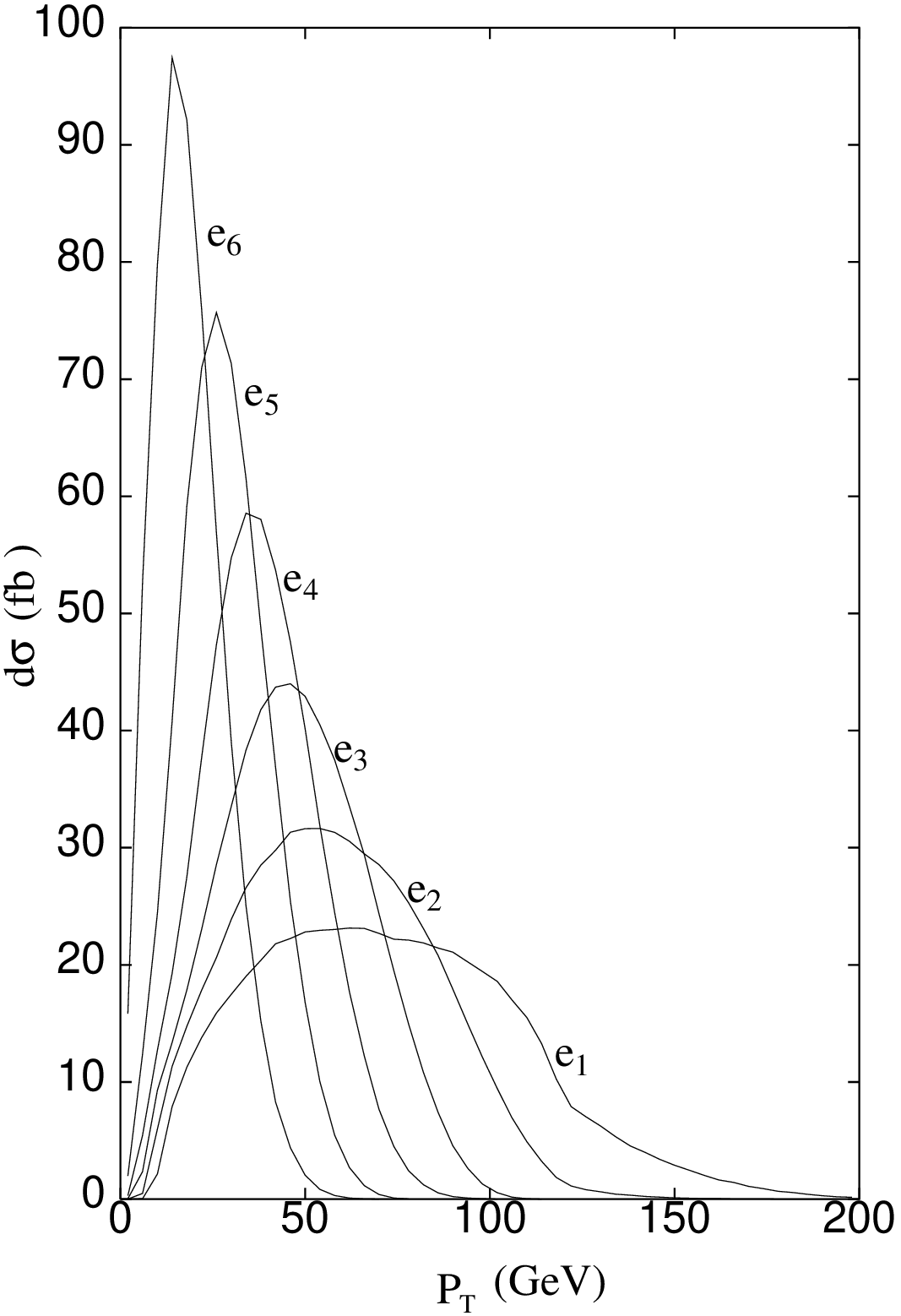,width=10cm}}
\vspace{.5in}
\caption{ $p_T$ distribution of leptons for the $\lambda_{ijk}$ case. The
MSSM input parameters are $\mu = -450$~(GeV), $M_2=200$~(GeV), and 
$\tan\beta = 2$. The number adjacent to each curve represents leptons 
with the following energy ordering: $E_{\ell_1}>E_{\ell_2}> E_{\ell_3} > 
E_{\ell_4}>E_{\ell_5}> E_{\ell_6} $.} 
\label{ptlep-lam}
\end{figure}

In Fig.~\ref{ptlep-lam}, we have shown the transverse momentum ($p_T$) 
distribution of the charged leptons produced in the final state for 
$M_{\er}=~150$ GeV, and the following the set of input parameters 
$\mu = -450 $~GeV, $M_2 = 200 $~GeV and $\tan\beta = 2$. 
For this set of parameter points $M_{\N0_1}=103$~GeV and $M_{\N0_2}=206$~GeV. 
For later studies of the distributions we will use this set of input 
parameters. It is easy to see from this distribution that all six leptons 
survive the $p_T^\ell> 5 $ GeV cut. Out of six leptons, two come from the 
decay of $\er$; the remaining four leptons come from the decay of $\N0_1$.  
 
We display in Table 1, some representative values of the cross 
sections in order to get an idea about the strength of the signal.  
In obtaining these numbers, we required six leptons satisfying the
criteria given in Eq.~(5), and in addition imposed that $\mpT > 15$~GeV.
The $\mpT$ requirement ensures that the signal is SM background free.
Two values of the right selectron mass, namely, 
$m_{{\tilde e}_R} = 150$~(GeV) and $m_{{\tilde e}_R} = 200$~(GeV), have been 
considered for the calculation of the cross sections. We have 
considered the actual branching ratios of the decays of right
selectrons including the direct decays through $R$-parity-violating
couplings as well as decays into heavier neutralinos. 
It has already been mentioned that if the decays into heavier neutralino 
states are allowed kinematically, they will lead to more complex signals 
which we have not considered in this work. For a fixed value of 
$\mu$ and $\tan\beta$, with increasing $M_2$, the LSP mass increases; 
hence the MSSM decay of $\er$ decreases because of phase-space suppression,
favoring the direct decay of $\er $ through $R$-parity-violating 
$\lambda_{231} $, which in turn reduces our signal. As is evident from 
this table, large cross sections may be obtained for a considerable region 
of the parameter space and with a projected integrated luminosity of 
50 ${\rm fb}^{-1}$ at an $e^- e^-$ collider one could see some 
thousands of events. It must be noted at this point that if taus 
are produced in the final state, they would decay mainly into hadrons, 
but that requires a separate analysis.  

\subsection{Signals from $\lambda'$-type couplings}
The decay pattern of the LSP changes as we go on to the $R$-parity-violating 
couplings of the type $\lambda' ~LQD^c$.
For example, $\lambda^{\prime}_{123}$ coupling gives
\bea
\N0_1 \goes \nu_e s \bar b,~e^- c b,~\bar\nu_e \bar s b,~e^+\bar c b. 
\eea
As before, we again consider a common value for all $\lambda'$-type
couplings. To identify the final state flavors one has to take into account 
the reduction in cross section due to flavor tagging efficiency. It
should be mentioned at this point that unlike the $\lambda$ case here all 
final states are not equiprobable.
We categorize the signals in the following manner. All these
states are assumed to be accompanied by two like-sign dielectrons arising
from $\er$ decay:
(1) $2\ell^\pm$ + jets; both $\N0_1\goes \ell^\pm jj$;
(2) jets + $\mpT$ ; both $\N0_1\goes \nu jj$;
(3) $\ell^\pm +{\rm jets} + \mpT$ ; one $\N0_1\goes \ell^\pm  jj$, the 
other $\N0_1\goes \nu jj$.

\begin{figure}[hbt]
\centerline{\epsfig{file= 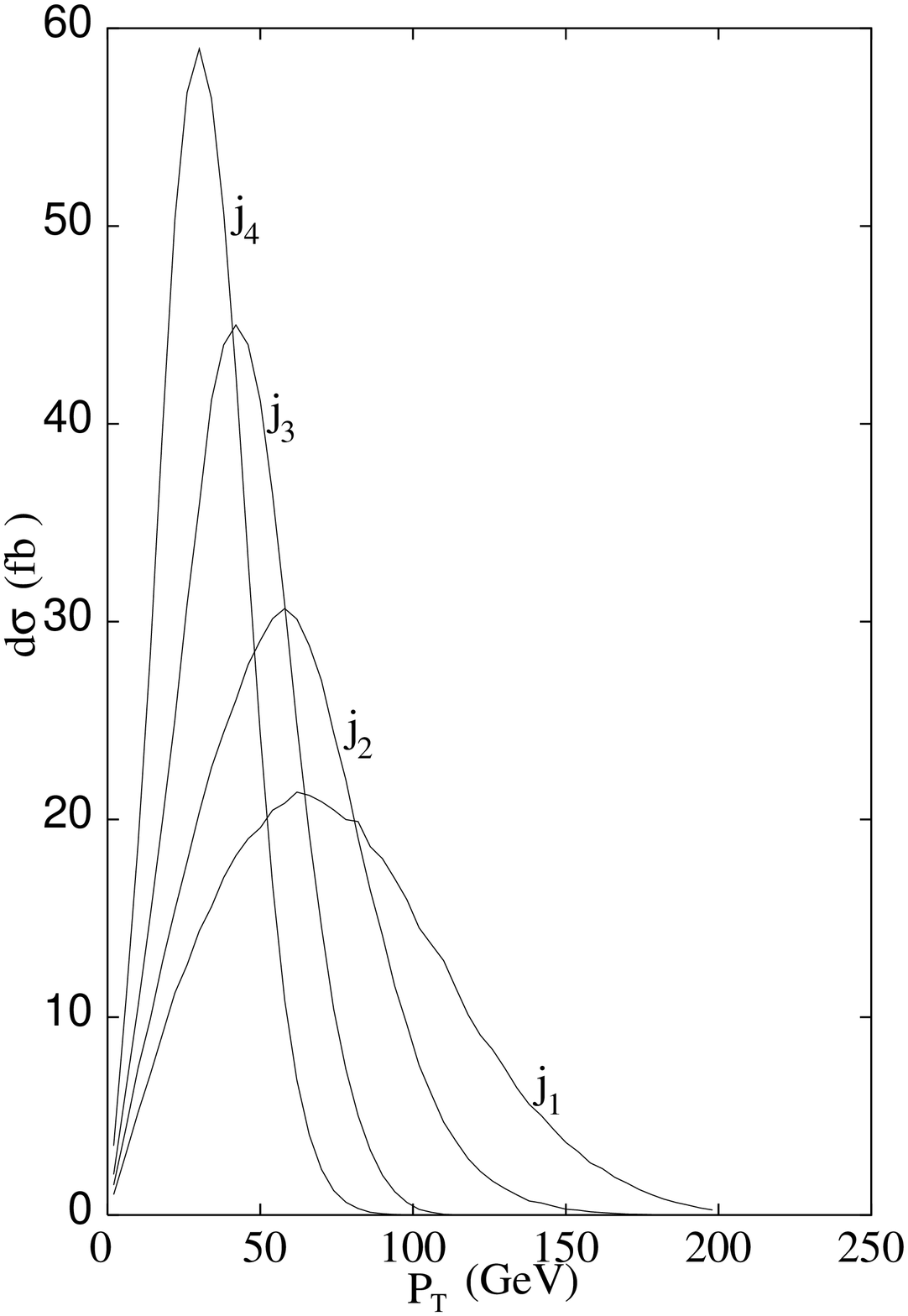,width=10cm}}
\vspace{1.0in}
\caption{ $p_T$ distribution of jets for both $\N0_1 \go \ell^\pm jj $
channels through $\lambda^\prime_{ijk}$ coupling. 
The number adjacent to each curve represents jets 
with the following energy ordering: $E_{j_1}>E_{j_2}> E_{j_3} > E_{j_4} $. 
The input parameters are same as in Fig.~\ref{ptlep-lam}.} 
\label{ptjet-lamp}
\end{figure}

\begin{figure}[hbt]
\centerline{\epsfig{file= 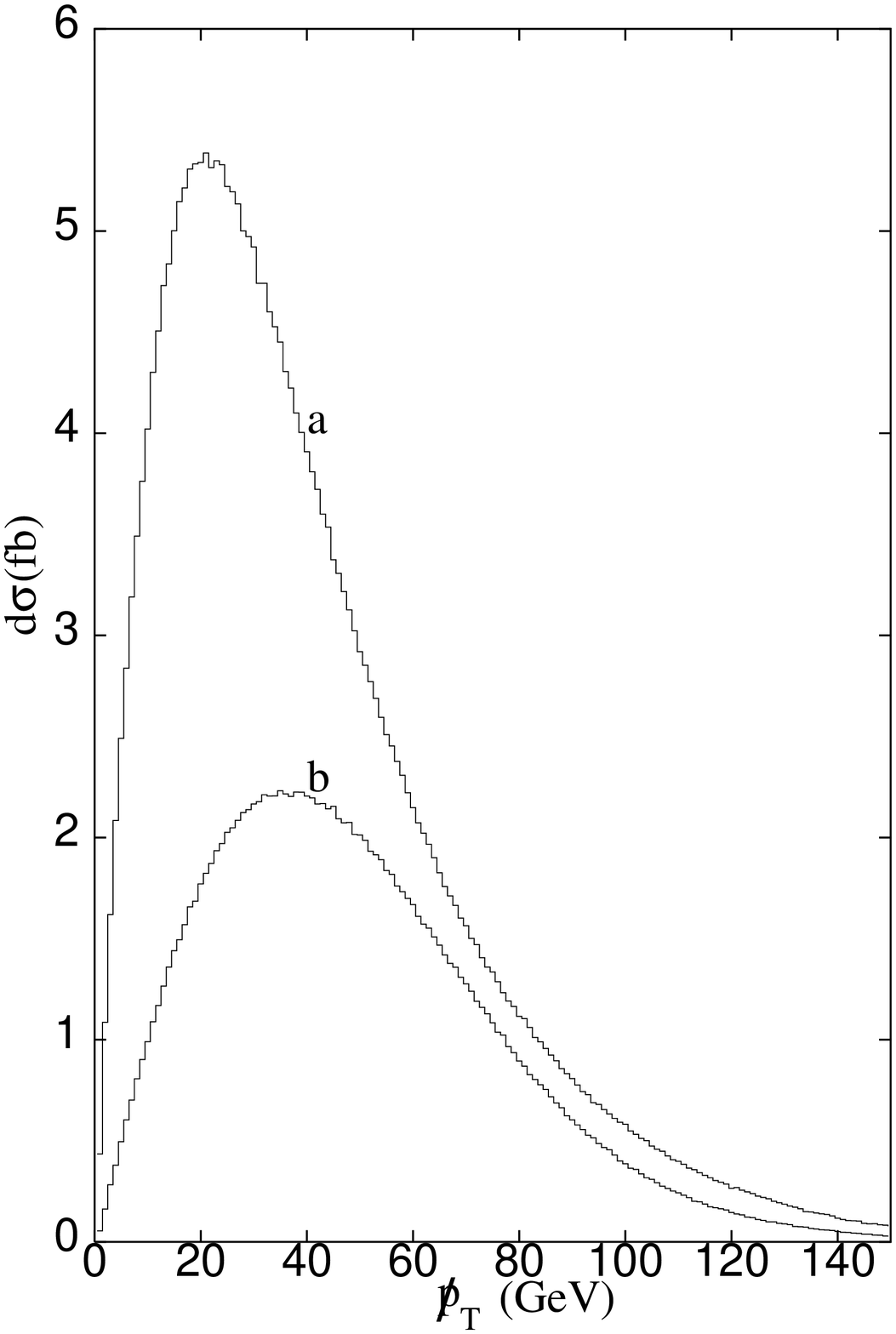,width=10cm}}
\vspace{1.0in}
\caption{ $\mpT$ distribution for the $\lambda^{\prime}_{ijk}$ case. 
$a$: both $\N0_1\goes \nu jj $.~$b$: one $\N0_1\goes \ell^\pm jj $, the 
other one $\N0_1\goes \nu jj $. The input parameters are same as in 
Fig.~\ref{ptlep-lam}.}
\label{mpt-lamp}
\end{figure}

The last channel will be enhanced by a combinatoric factor of
2. We have folded the cross section with the branching fraction of 
the LSP.  
The selection cuts (as discussed earlier) are applied to the leptons 
and jets. After the energy ordering ($E_{j_1}>E_{j_2}>E_{j_3} >E_{j_4}$), 
we study the jet $p_T$ distribution as shown in Fig.~\ref{ptjet-lamp}. 
These jets and charged leptons are also associated with large $\mpT$
arising from neutrinos for channels ($2$) and ($3$) listed above. 
The $\mpT$ distribution is shown in 
Fig. \ref{mpt-lamp}. The distribution ${\it a}$ corresponds to the 
case when both the LSP decays into the $\nu jj$ channel, where as ${\it
b}$ 
represents the $\mpT$ distribution when one of the LSP decays through 
the $\nu jj$ mode and the other one through the $\ell^\pm jj$ mode.
 
Finally in Table 2 we give cross sections for signals for two
$\er$ masses 150 GeV and 200 GeV.
We required four jets and, respectively, four, three, or two leptons 
satisfying the criteria given in Eqs.~(6) and (5), for channels ($1$),
($2$), and ($3$) 
listed above. In addition, $\mpT > 15$~GeV is imposed for channels ($2$) and
($3$). Cross sections for heavier right-selectron
mass ($=200$~GeV) are lower than the corresponding quantities for 150 GeV 
$\er$ mass, just because of a lack of
enough phase space. The difference in the three cross sections in each row can
be explained from the branching ratio of $\N0_1$ in two different channels
$\ell^\pm jj$ and $\nu jj$. The inputs remain same as in Table 1.
The cross sections for these various channels are fairly large over a 
wide region of parameter space which is accessible 
in a 500 GeV $e^- e^-$ collider. Signals corresponding to 
$e^- e^- + {\rm jets} + \mpT$ and 
$e^- e^- + \ell^\pm + {\rm jets} + \mpT$ final states may have the 
standard model background coming from $W^- W^- Z Z$ production. But 
this cross section is found to be too low ($< 40 $~fb) and does not 
affect the signal in a significant way. 

\begin{figure}[hbt]
\centerline{\epsfig{file= 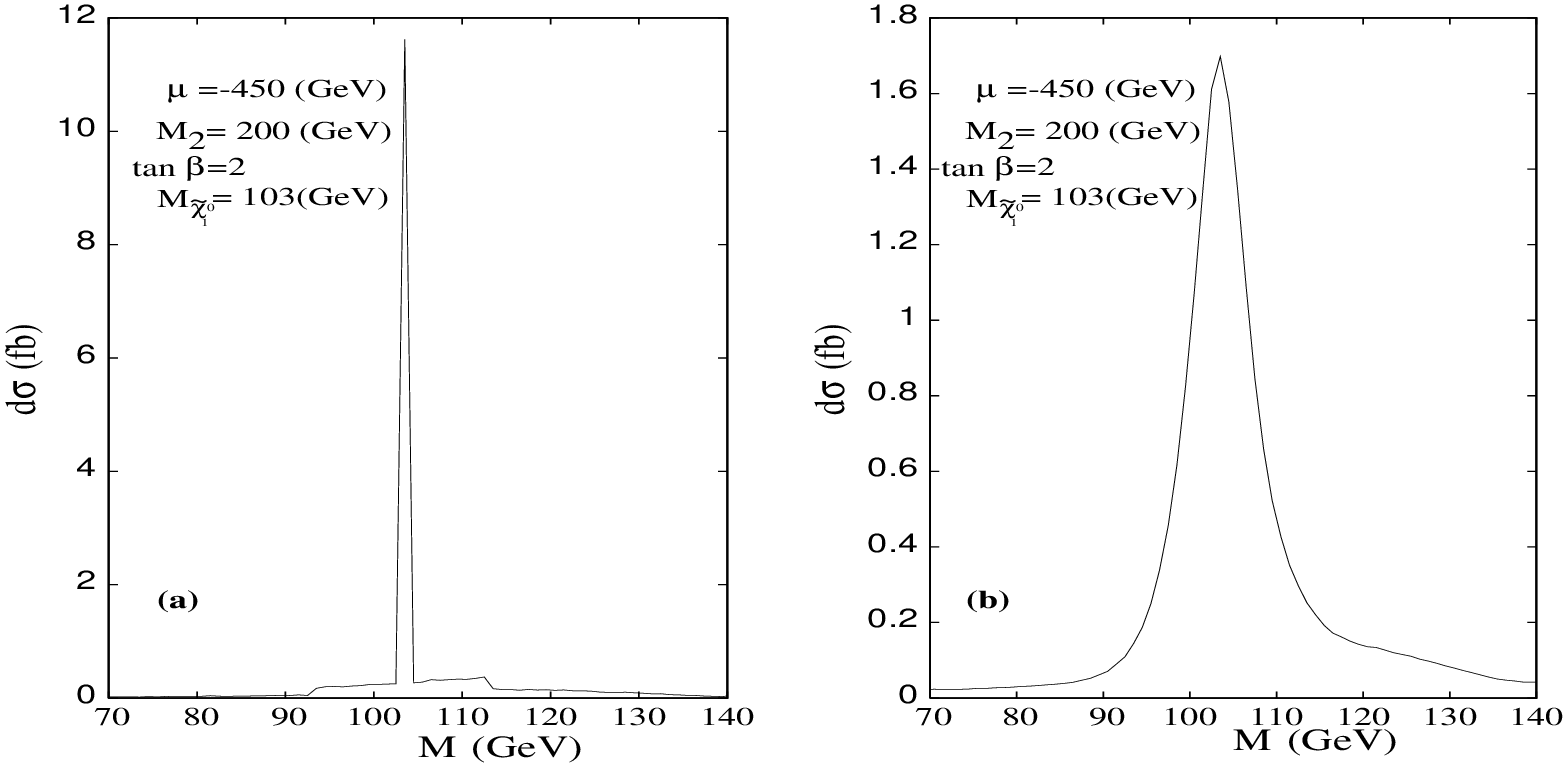,width=10cm}}
\vspace{.5in}
\caption{ Distribution in invariant mass reconstruction from the lepton
and all jets in the same hemisphere. (a) without lepton and jet energy 
smearing and (b) with lepton and jet energy smearing.
Both $\N0_1$ decay into the $\ell^\pm jj $ channel through 
$\lambda^\prime_{ijk}$ coupling. } 
\label{invmass-lamp}
\end{figure}

If the produced LSP is highly relativistic, then its decay products will 
be confined within a narrow cone around the direction of the LSP. In that 
case, the lepton (decaying from LSP) in a particular hemisphere is identified 
and its invariant mass is constructed with all jets in the 
same hemisphere. A similar thing is done in the opposite hemisphere. Then 
we demand that these two invariant masses should lie within 10 GeV of each 
other. If these two invariant masses are equal or nearly equal, we can say 
that they arise from the same parent particle. In
Fig.~\ref{invmass-lamp}(a) 
we represent such an invariant mass distribution, which shows a distinct peak 
at the LSP mass ($=103$~GeV). In order to get an estimate of the mass
resolution, we have used Gaussian smearing~\cite{peskin-murayama} of the 
energies of jets and leptons to ``mimic" the response of a detector:  
\bea
\Delta E_j/E_j = 0.4/\sqrt{E_j} + 0.02,~~~~~~\Delta E_l/E_l = 0.15/\sqrt{E_l}
+ 0.01.
\eea
In Fig. \ref{invmass-lamp}(b), we show the mass distribution after
energy smearing. The LSP mass determined in this way has resolution 
$\Delta M/M = 4\%$. It should also be noted that for about 80$\%$ of 
the total events the mass reconstructed in both sides lies within 10 GeV 
of each other. In principle one can also reconstruct the selectron mass 
in this way, but a more precise determination can be done by 
a threshold scan~\cite{feng}.  

\subsection{Signals from $\lambda^{\prime\prime}$-type couplings}
\begin{figure}[hbt]
\centerline{\epsfig{file= 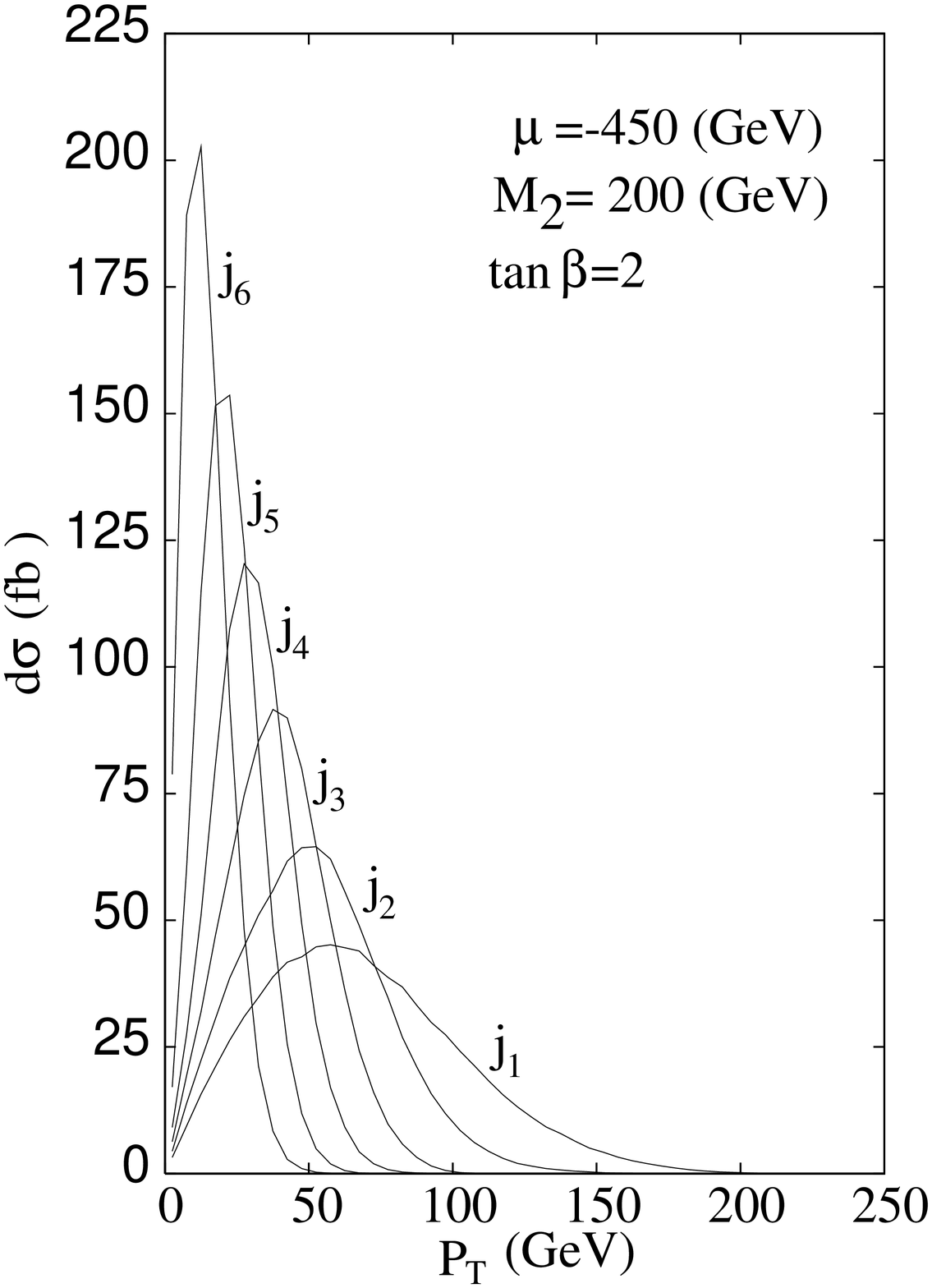,width=10cm}}
\vspace{1.0in}
\caption{ $p_T$ distribution of jets for the $\lambda^{\prime\prime}_{ijk}$ 
case. The number adjacent to each curve represents jets 
with the following energy ordering: $E_{j_1}>E_{j_2}> E_{j_3} > E_{j_4}> 
E_{j_5}>E_{j_6}$.} 
\label{ptjet-lpp}
\end{figure}

\begin{figure}[hbt]
\centerline{\epsfig{file= 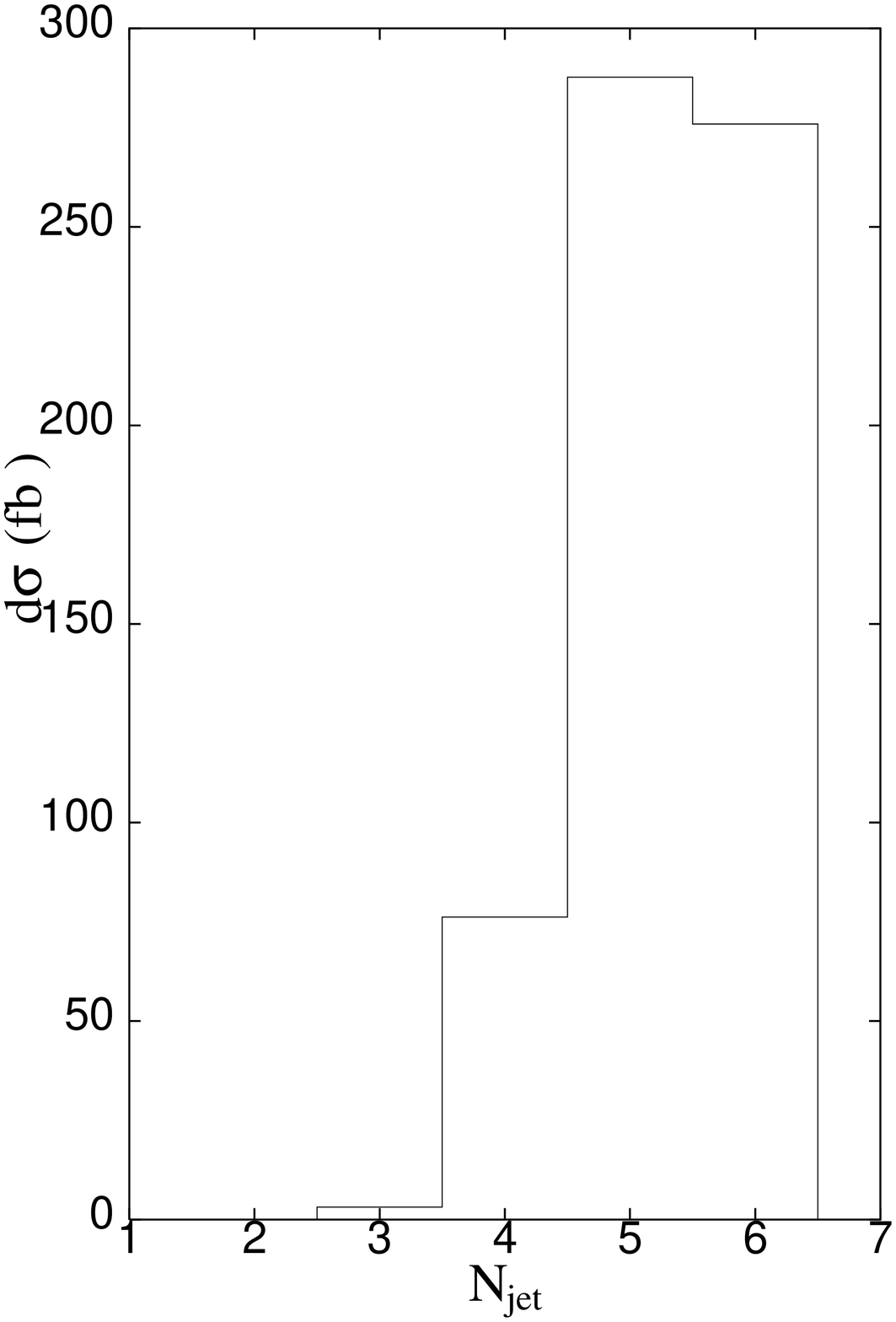,width=10cm}}
\vspace{1.0in}
\caption{ Number of jets ($N_{jet}$) when $\N0_1\go jjj $ through 
$\lambda^{\prime\prime}_{ijk}$ coupling.}  
\label{njet-lpp}
\end{figure}

Finally, the presence of $\lambda^{\prime\prime}$ in the superpotential can
induce $B$ number violating decay of the LSP. In this case, the LSP will 
simply decay into three hadronic jets: 
\bea
e^- e^- \goes \er^- + \er^- \go e^- + e^- + \N0_1 + \N0_1 \go e^- e^-
+6~{\rm jets}, 
\eea
where the sets of three jets have invariant mass peaking at the neutralino 
mass (assuming all jets are seen). As before, we impose the selection cuts
on leptons and at least four jets. From the $p_T$ distribution of six jets in 
Fig.~\ref{ptjet-lpp} it is clear that for this value of the LSP mass 
($=103$~GeV), most of the jets are hard enough to satisfy the jet trigger 
requirement as discussed previously. From the jet number distributions in
Fig.~\ref{njet-lpp}, we see that most of the time the cross section prefers 
to peak at the five-jet channel ($44.69\%$ of the events), followed by
the six-jet ($42.86\%$ of the events) and four-jet ($11.83\%$ of the events) 
channels. 
The three-jet fraction of the cross section is less than $0.5\%$. Imposition 
of $p^j_T> 15 $~GeV and $|\eta_j|< 3$ cuts on the jets reduces the jet number. 
Finally we also merge two jets into a single jet if their angular 
separation $\Delta R_{jj}< 0.7 $. 
The probability of jet merging is highly dependent on the mass of the 
parent particle from which the jets originate and also on $\sqrt{s}$.
The larger the boost of the parent particle, the higher the probability of 
jet merging.  
In this case, a 103 GeV LSP is produced from the decay of a 150 GeV right 
selectron. Each of these LSPs then decays into three jets with a reasonable 
boost, leading six jets to merge into five jets and occasionally into
four and three jets.  

In Table 3 we give signal cross sections for some representative values of
parameters. In this case, we assume the squark mass to be 500 GeV, which
enters as a propagator in the decay LSP. One can also reconstruct the LSP mass
using the following strategy: selecting the hardest jet in the final state, 
its invariant mass is then constructed with all other jets in that hemisphere. 
A similar thing is done in the opposite hemisphere. Then we demand that these
two invariant masses should lie within 10 GeV of each other. If these two
invariant masses are equal or nearly equal, we can say that they arise from
the same parent particle. Though we will not present here the invariant mass
distribution, similar kinds of studies have been done by other authors and
also by the ALEPH Collaboration in their study of the (now defunct) four-jet 
anomaly~\cite{Ghosh1,Ghosh2,ALEPH_4jet}. 

Before we conclude, we would like to mention the possible SM backgrounds
in this case. We have earlier found that most of 
the time the signal cross section prefers to peak around five and six jets, 
free from any SM background. However, there is a small fraction of 
cross section that goes into four-jet channels, which is less important for 
our purpose as far as the SM backgrounds are concerned. This particular 
signal has SM backgrounds from 
(a) $e^-e^-\go e^-e^- ZZ$, (b) $e^-e^-\go e^-e^- Z^\ast Z$, and (c) 
$e^-e^-\go e^-e^- Z^\ast Z^\ast$, with hadronic decay of $Z$ (assuming all 
jets are seen). 

One can make a rough estimate for this background. After putting the
selection cuts and including the relevant branching ratios the cross section
for $e^-e^-\go e^-e^-Z$ is of the order of $100$~fb. 
This cross section will get electroweak suppression if another $Z$ boson 
is radiated; moreover, the ${\rm Br.}(Z\go q\bar q)$ will further reduce this.
After all these, if this background is still comparable to the signal, 
then this can be eliminated by imposing the condition that the pair of
dijet invariant mass $M_{jj}$ should not peak around $M_Z$. However, 
this may reduce the signal cross section in the region of parameter 
space where the LSP mass is nearly degenerate with $M_Z$. The detailed 
calculation of the other two backgrounds [(b) and (c)] is very cumbersome 
and we will not perform this here. In this case, our main thrust will be to 
count the jets in the final state (associated with two electrons) 
to distinguish it from the background.

\section{Conclusions}
We have discussed the pair production of right selectrons at a 500 GeV
$e^-e^-$ linear collider in the $R$-parity-violating supersymmetric
model. The decay of right selectrons can yield a final state with an 
electron and a neutralino, mostly the LSP. Hence, we have two like-sign 
dielectrons and neutralinos in the final state. We have assumed that 
$R$ parity is weakly violated and thus only the LSP will decay into 
multifermion states. Different possibilities have been considered and it 
seems that rather optimistic signals can be seen for this kind of model.
The decay of the LSP gives charged leptons, jets, and neutrinos 
in the final state. The behavior of these leptons, jets, and missing
transverse momentum (mainly due to neutrinos) has 
been analyzed using a parton level Monte Carlo event generator. This also
enables us to study the approximate distributions for different kinematic
variables of leptons and jets. 
The decay of the LSP through $L$-number-violating coupling~($\lambda$) leads 
to a very distinct signal with hard isolated leptons and large 
missing transverse momentum. There are no SM processes which can mimic 
this signal. Similarly, for $\lambda^\prime_{ijk}$ couplings,
the signal basically consists of charged leptons, multiple jets, and/or
missing transverse momentum. In addition to this, the Majorana nature of
the LSP gives rise to like-sign dilepton signals with practically no SM 
backgrounds.
It has been demonstrated that the reconstruction of the lepton-jet invariant 
mass can  give a rough estimate for the LSP mass. For 
$\lambda^{\prime\prime}_{ijk}$ coupling, the final state will have multiple
jets associated with like-sign dielectrons. 
We have shown that proper jet counting is required to distinguish the
signal from the SM backgrounds. In this case also it might be possible 
to determine the LSP mass from the jet invariant mass reconstruction. 

\vspace{0.52in}
\begin{center}
{\bf Acknowledgments}
\end{center}
The authors are grateful to Biswarup Mukhopadhyaya and Sreerup Raychaudhuri
for helpful discussions.
\newpage
\footnotesize

\def\pr#1,#2 #3 { {\em Phys.~Rev.}        ~{\bf #1},  #2 (19#3) }
\def\prd#1,#2 #3{ {\em Phys.~Rev.}       ~{D \bf #1}, #2 (19#3) }
\def\pprd#1,#2 #3{ {\em Phys.~Rev.}      ~{D \bf #1}, #2 (20#3) }
\def\prl#1,#2 #3{ {\em Phys.~Rev.~Lett.}  ~{\bf #1},  #2 (19#3) }
\def\plb#1,#2 #3{ {\em Phys.~Lett.}       ~{\bf B#1}, #2 (19#3) }
\def\npb#1,#2 #3{ {\em Nucl.~Phys.}       ~{\bf B#1}, #2 (19#3) }
\def\prp#1,#2 #3{ {\em Phys.~Rep.}       ~{\bf #1},  #2 (19#3) }
\def\zpc#1,#2 #3{ {\em Z.~Phys.}          ~{\bf C#1}, #2 (19#3) }
\def\epj#1,#2 #3{ {\em Eur.~Phys.~J.}     ~{\bf C#1}, #2 (19#3) }
\def\mpl#1,#2 #3{ {\em Mod.~Phys.~Lett.}  ~{\bf A#1}, #2 (19#3) }
\def\ijmp#1,#2 #3{{\em Int.~J.~Mod.~Phys.}~{\bf A#1}, #2 (19#3) }
\def\ptp#1,#2 #3{ {\em Prog.~Theor.~Phys.}~{\bf #1},  #2 (19#3) }

\newpage
\vspace*{2.0in}
 \input{serser.fig}

\newpage

\newpage
\begin{table}[b]
\begin{center}
\begin{tabular}{|c|c|c|c|c|c|} \hline
\multicolumn{4}{|c|}{} & {$m_{{\tilde e}_R} =
150$~(GeV)} & {$m_{{\tilde e}_R} = 200$~(GeV)} \\ \hline
$\mu$~(GeV) & $M_2$~(GeV) & $\tan\beta$ & $m_{\tilde \chi^0_1}$~(GeV) &
{$\sigma$~(fb)} & {$\sigma$~(fb)} \\ \hline
-450 & 200 & 2 & 103.1 & 537.40  & 407.0    \\ \hline
-375 & 250 & 2 & 128.4 & 163.22  & 380.86   \\ \hline
 400 & 250 & 2 & 118.9 & 334.86  & 410.11   \\ \hline
-500 & 280 & 20& 140.2 & 11.93   & 325.05   \\ \hline
 375 & 200 & 20& 98.5  & 564.31  & 398.30   \\ \hline
 475 & 265 & 20& 131.6 & 107.16  & 372.35   \\ \hline
-480 & 300 & 40& 149.8 & 0.0     & 252.30   \\ \hline
-350 & 225 & 40& 111.5 & 448.23  & 413.10   \\ \hline
 400 & 200 & 40&  99.0 & 563.24  & 400.41   \\ \hline
\end{tabular}
\end{center}
\caption {Signal ($ e^- e^- + 4\ell^\pm  + \mpT $ ) cross section
          assuming LSP decays through $\lambda_{ij1}$ coupling
          for some representative values of the input parameters. For 
          $\lambda_{ijk}$ with $k\neq 1$, these cross sections would be larger
          by at least a factor of $2$ since, in that case, the ${\tilde e}_R$ 
          only has $R$-parity-conserving decay modes.}
\end{table}

\newpage

\newpage

\begin{table}[b]
\begin{center}
\begin{tabular}{|c|c|c|c|c|c|} \hline
\multicolumn{4}{|c|}{} & {$m_{{\tilde e}_R} =
150$~(GeV)} & {$m_{{\tilde e}_R} = 200$~(GeV)} \\ \hline
$\mu$~(GeV) & $M_2$~(GeV) & $\tan\beta$ & $m_{\tilde \chi^0_1}$~(GeV) &
{$\sigma$~(fb)} & {$\sigma$~(fb)} \\ \hline
     &     &   &       &184.40  & 130.01  \\
-450 & 200 & 2 & 103.1 &214.54  & 150.20   \\
     &     &   &       &126.85  & 89.43  \\ \hline
     &     &   &       &234.20  & 180.47  \\
-375 & 250 & 2 & 128.4 &322.0   & 245.66   \\
     &     &   &       &180.80  & 137.70  \\\hline
     &     &   &       &58.68   & 43.46  \\
 400 & 250 & 2 & 118.9 &239.80  & 175.72  \\
     &     &   &       &137.00  & 100.50 \\\hline
     &     &   &       &120.26  & 125.81 \\
-500 & 280 & 20& 140.2 &277.60  & 289.10   \\
     &     &   &       &152.91  & 158.83  \\\hline
     &     &   &       &73.95   & 51.71 \\
 375 & 200 & 20& 98.5  &184.42  & 128.8    \\
     &     &   &       &110.64  & 77.31 \\ \hline
     &     &   &       &126.84  & 100.50   \\
 475 & 265 &20 & 131.6 &321.33  & 253.00   \\
      &     &   &      &179.23  & 140.38   \\\hline
      &     &   &      &2.79    & 130.48      \\
 -480 & 300 & 40& 149.8 &7.23   & 320.20     \\
      &     &   &       &4.25   & 173.53     \\\hline
      &     &   &       &101.20 & 72.85       \\
 -350 & 225 & 40& 111.5 &246.40 & 176.07    \\
     &     &   &       &142.30  & 102.0      \\\hline
     &     &   &       &82.48   & 57.60     \\
 400 & 200 & 40& 99.0  &190.30  & 132.70    \\
     &     &   &       &114.06  & 79.80      \\\hline
\end{tabular}
\end{center}
\caption {Signal cross section assuming LSP decays through
          $\lambda^{\prime}_{ijk}$ coupling for some representative
          values of the input parameters. In each row, the first,
          second, and third numbers correspond to cross sections for
          the following final states: $e^-e^- +  2\ell + {\rm jets} $,
$e^-e^- +\ell + \mpT + \rm jets $, and $e^-e^- + {\rm jets} + \mpT $,
respectively.}
\end{table}

\newpage

\newpage
 
\begin{table}[hbt]
\begin{center}
\begin{tabular}{|c|c|c|c|c|c|} \hline
\multicolumn{4}{|c|}{} & {$m_{{\tilde e}_R} =
150$~(GeV)} & {$m_{{\tilde e}_R} = 200$~(GeV)} \\ \hline
$\mu$~(GeV) & $M_2$~(GeV) & $\tan\beta$ & $m_{\tilde \chi^0_1}$~(GeV) &
{$\sigma$~(fb)} & {$\sigma$~(fb)} \\ \hline
-450 & 200 & 2 & 103.1 &645.65  & 447.14   \\ \hline
-375 & 250 & 2 & 128.4 &925.61  & 714.22\\ \hline
 400 & 250 & 2 & 118.9 &859.80  &634.50  \\ \hline
-500 & 280 & 20& 140.2 &803.53  & 842.07 \\ \hline
 375 & 200 & 20& 98.5  &588.11  & 404.50 \\ \hline
 475 & 265 & 20& 131.6 &953.10  & 758.86 \\ \hline
-480 & 300 & 40& 149.8 & 45.0   & 929.10  \\ \hline
-350 & 225 & 40& 111.5 &758.20  & 540.92    \\ \hline
 400 & 200 & 40& 99.0  &595.21  & 409.54  \\ \hline

\end{tabular}

\end{center}
\caption {Signal ($ e^- e^- +{\rm jets} $ ) cross section
          assuming LSP decays through $\lambda^{\prime \prime}_{ijk}$
          coupling for some representative values of the input
          parameters.}
\end{table}

\end{document}